\documentclass[twocolumn,showpacs,preprintnumbers,amsmath,amssymb,superscriptaddress]{revtex4-2} 
\usepackage{chemformula} 
\usepackage{graphicx}
\usepackage[colorlinks=true,citecolor=blue,urlcolor=blue,linkcolor=blue,bookmarksopen]{hyperref}
\usepackage{color}
\usepackage{braket}
\usepackage{physics}

\begin{document}

\title{Simultaneous observation of bright and dark polariton states in subwavelength gratings made from quasi-bulk WS$_2$}

\author{Paul Bouteyre}
\email{p.bouteyre@sheffield.ac.uk}
\affiliation{Department of Physics and Astronomy, University of Sheffield, Sheffield S3 7RH, U.K} 
\author{Xuerong Hu}
\affiliation{Department of Physics and Astronomy, University of Sheffield, Sheffield S3 7RH, U.K} 
\author{Sam A. Randerson}
\affiliation{Department of Physics and Astronomy, University of Sheffield, Sheffield S3 7RH, U.K} 
\author{Panaiot G. Zotev}
\affiliation{Department of Physics and Astronomy, University of Sheffield, Sheffield S3 7RH, U.K} 
\author{Yue Wang}
\affiliation{School of Physics, Engineering and Technology, University of York, York, YO10 5DD, UK} 
\author{Alexander I. Tartakovskii}
\email{a.tartakovskii@sheffield.ac.uk}
\affiliation{Department of Physics and Astronomy, University of Sheffield, Sheffield S3 7RH, U.K} 

\date{\today}

\begin{abstract}

Over the last decade, layered crystals, dubbed van der Waals (vdW) materials, have attracted tremendous interest due to their unique properties in their single and few layer regimes. Their bulk counterparts, however, have only been recently explored as building blocks for nanophotonics as they offer promising properties such as high refractive indices and adherence to any type of substrates. We present here a variety of 1D grating structures composed of bulk transition metal dichalcogenide (TMD) WS$_2$ as a highly tunable and versatile platform for observation of multi-level polaritonic system. The WS$_2$ excitons are simultaneously strongly coupled with the two grating photonic modes including the Bound State in the Continuum (BIC) of the lower energetic mode giving rise to polariton-BICs (pol-BICs). The polaritonic dispersion shapes can be varied in a straightforward fashion by choosing WS$_2$ films of different thicknesses and by changing the period of the grating. 

\end{abstract}


\maketitle

Exciton-polaritons \cite{Weisbuch1992}, half-light half-matter quasiparticles, arise from the strong coupling regime between an excitonic and a photonic state when the coupling rate between the two states is faster than the respective dissipation rates. Exciton-polaritons have been demonstrated in many nanofabricated photonic strutures such as microcavities \cite{Weisbuch1992,Kasprzak2006,Landau2022}, photonic crystals waveguides \cite{Maggiolini2023,Ardizzone2022,Riminucci2023,Hu2022,Khestanova2024,Wu2024,Dang2022,Kim2021,Wang2023,Zhang2020,Zong2021,Kravtsov2020} or micropillar lattices \cite{Gagel2024,StJean2017,Kuriakose2022}. From their hybrid nature, the exciton-polaritons hold the properties of their light and matter counterparts, and therefore are nonlinear bosons that can ballistically propagate. Using these properties, many proofs-of-concept of polaritonic devices, potential candidate for future all-optical devices, were demonstrated \cite{Kasprzak2006,Gao2012,Nguyen2013,Sturm2014,Gao2012,Marsault2015,Lee2024}. Moreover, thanks to the excitonic part of polaritons, several nonlinear effects were reported \cite{Maggiolini2023,Ardizzone2022,Riminucci2023,StJean2017,Kravtsov2020,Kuriakose2022,Wu2024,Khestanova2024,Gagel2024}, such as polaritonic lasing \cite{Ardizzone2022,Riminucci2023,Gagel2024,StJean2017,Landau2022,Wu2024} or few-photon all-optical phase rotation \cite{Kuriakose2022}.

Topological properties of exciton-polaritons were also explored using topological photonic structures \cite{Gagel2024,StJean2017,Kim2021,An2024}, as well as non-Hermitian properties through the strong coupling between semiconductor excitons and Bound States in the Continuum (BIC) in photonic structures giving rise to the so called polariton-BICs \cite{Wu2024,Hu2022,Riminucci2023,Ardizzone2022,Maggiolini2023,Dang2022,Wang2023,Zong2021,Kravtsov2020,Kim2021}. The BICs \cite{VonNeumann1929,Friedrich1985}, intriguing non-Hermitian properties \cite{ElGanainy2019}, are dark photonic states forbidden to couple to the continuum due to destructive interferences between lossy photonic modes in symmetrical photonic structures.

In the meantime, layered crystals, dubbed van der Waals (vdW) materials, have over the last two decades attracted tremendous interest due to their unique properties in their single and few layer regimes. Indeed monolayers of diverse 2D materials, especially of transition metal dichalcogenides (TMDs), possess strong excitonic properties and quantum yields much higher than well-studied III-V quantum wells at room temperature. For these reasons, monolayers of vdWs materials have been integrated in various silicon, silicon nitride, or III-V semiconductor based photonic structures ranging from polaritonic devices in microcavities \cite{Zhang2018}, to polaritons in Spin-Hall topological lattices \cite{Li2021}.

The bulk counterparts of these 2D materials, have been less explored for potential use in nanophotonics as the direct bandgap of TMDs becomes indirect in the multilayer form, which considerably reduces their quantum yields. However, bulk 2D materials offer very attractive properties for realizing photonic building blocks. Similarly to monolayers, they can be readily fabricated via mechanical exfoliation and they adhere easily to a wide range of substrates without need for chemical bonding or lattice matching. It was also recently shown that standard lithography can be used to pattern 2D materials leading to high quality structures \cite{Munkhbat2020,Isoniemi2024}. As bulk materials, TMDs have a variety of transmission windows with low losses, present uniaxial anisotropy and they have a higher refractive index versus bandgap than is expected from the Moss rule \cite{Moss1985,Zotev2023}. These properties make them ideal for slab waveguides. 

\begin{figure*}[!ht]
\centering
\includegraphics[width=\linewidth]{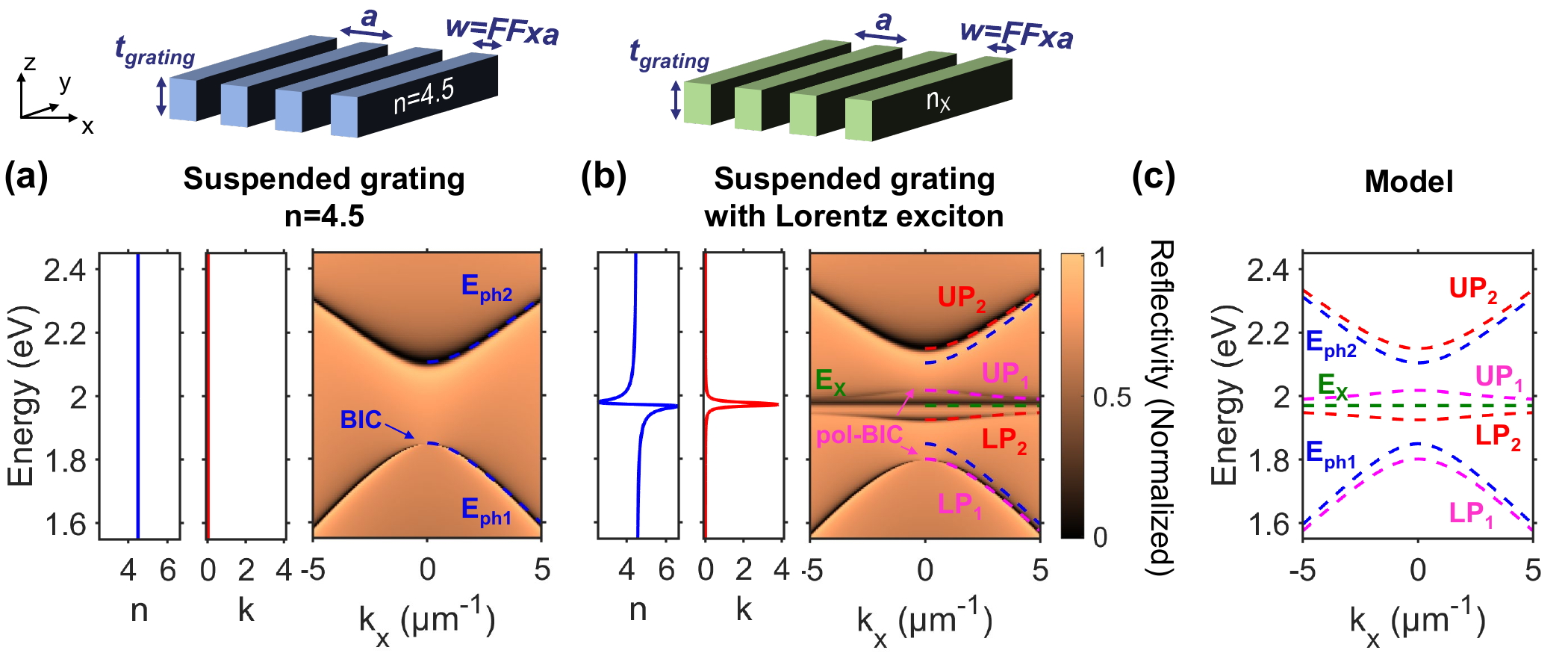}
  \caption{\textbf{Analytical model on simulated suspended gratings.} (a) Simulated reflectivity performed with Rigorous Coupled Wave Analysis (RCWA) of a suspended grating ($a$=360nm, $FF$=0.8 $t$= 20nm) with a constant refractive index with no losses (n=4.5, k=0). The two photonic modes of the grating, $E_{ph1}$ and $E_{ph2}$, are fitted with equation \ref{eq1} ($U$=255 meV, $E_0$=1.98 eV, $v_g^{ph1}$=72 meV.$\mu$m, $v_g^{ph2}$=62 meV.$\mu$m), and are depicted by blue dash lines. A Bound State in the Continuum (BIC) can be observed on the lower energetic photonic mode $E_{ph1}$. (b) RCWA simulated reflectivity of the same suspended grating structure but with a refractive index $n_X$ including an excitonic feature modelized with a Lorentz oscillator (more details on methods). Both photonic modes $E_{ph1}$ and $E_{ph2}$ (blue dashed lines)  described in (a), strongly couple with the exciton, creating two sets of lower ($LP$), and upper ($UP$) polaritonic dispersions: ($LP_{1}$, $UP_{1}$), pink dashed lines, and ($LP_{2}$, $UP_{2}$), red dashed lines. The BIC in (a) also strongly couples with the exciton giving rise to two polariton-BICs (pol-BICs) in pink. Some excitons ($E_X$ green dashed lines) remains uncoupled to the photonic modes ($E_{ph1}$,$E_{ph2}$) and can be seen as a horizontal green dashed line. (c) Polaritonic dispersions of (b) obtained using the photonic modes parameters in (a) and the model in equation \ref{eq2}, with $E_X=1.98$ eV and $V$=90 meV.}
  \label{figure1}
\end{figure*}

Exciton-polaritons were observed in bulk WS$_2$ photonic crystals in which the TMD served both as the excitonic material and building block of the photonic structure \cite{Zotev2023,Randerson2024,Munkhbat2019}. Indeed, bulk WS$_2$ has a direct local bandgap (as opposed to the global bandgap) at the K point and still exhibits strong excitons \cite{Zotev2023,Munkhbat2019}. Propagation of exciton-polaritons as well as hybrid exciton-plasmon-exciton were studied in bulk WS$_2$-based gratings \cite{Cho2023,Zhang2020}. However, the configurations of these structures were such that the non-hermiticity of the photonic system could not be explored such as the strong coupling between the WS$_2$ exciton and the grating BICs. Moreover, the WS$_2$ excitons were strongly coupled to only one of the grating modes. \\

In this context, we present here a highly tunable and versatile platform for observation of a double polaritonic system based on a bulk van der Waals excitonic material. We realize a variety of 1D grating structures composed of transition metal dichalcogenide WS$_2$ multilayers of thicknesses 10 to 60 nm on SiO$_2$ and Au substrates with respective grating periods of 350 nm and 400 nm and filling factors 0.55 and 0.7. The WS$_2$ excitons are simultaneously strongly coupled with the two grating photonic modes including the BIC of the lower energetic mode giving rise to polariton-BICs (pol-BICs). The polaritonic dispersions shapes can be varied in a straightforward fashion by choosing WS$_2$ films of different thicknesses and by changing the period of the grating.

The combination of the various advantages of 2D materials and of polaritons in a highly tunable and versatile platform, opens the way to a new prospect of future polaritonic photonic devices. Indeed, it was proposed that entire photonic circuits could be made of TMDs \cite{Ling2021}. Moreover, the highly tunable subwavelength gratings presented in our work could also be used in conjunction with slabs and whole photonic structures made from van der Waals or many other materials such as III-V membranes placed on top, thus opening the way to 3D nanophotonic heterointegration and straightforward fabrication of sophisticated polaritonic devices. 


In the following, we will first describe the double polaritonic system by performing RCWA (Rigorous Coupled Wave Analysis) simulations of a suspended grating of refractive index close to that of WS$_2$. Finally, we will experimentally demonstrate the double polaritonic system in WS$_2$-based gratings on SiO$_2$/Si and Au substrates in angle-resolved reflectivity measurements.

\vspace*{-10pt}

\section{Model and numerical results}

\begin{figure*}[ht!]
\centering
  \includegraphics[width=\linewidth]{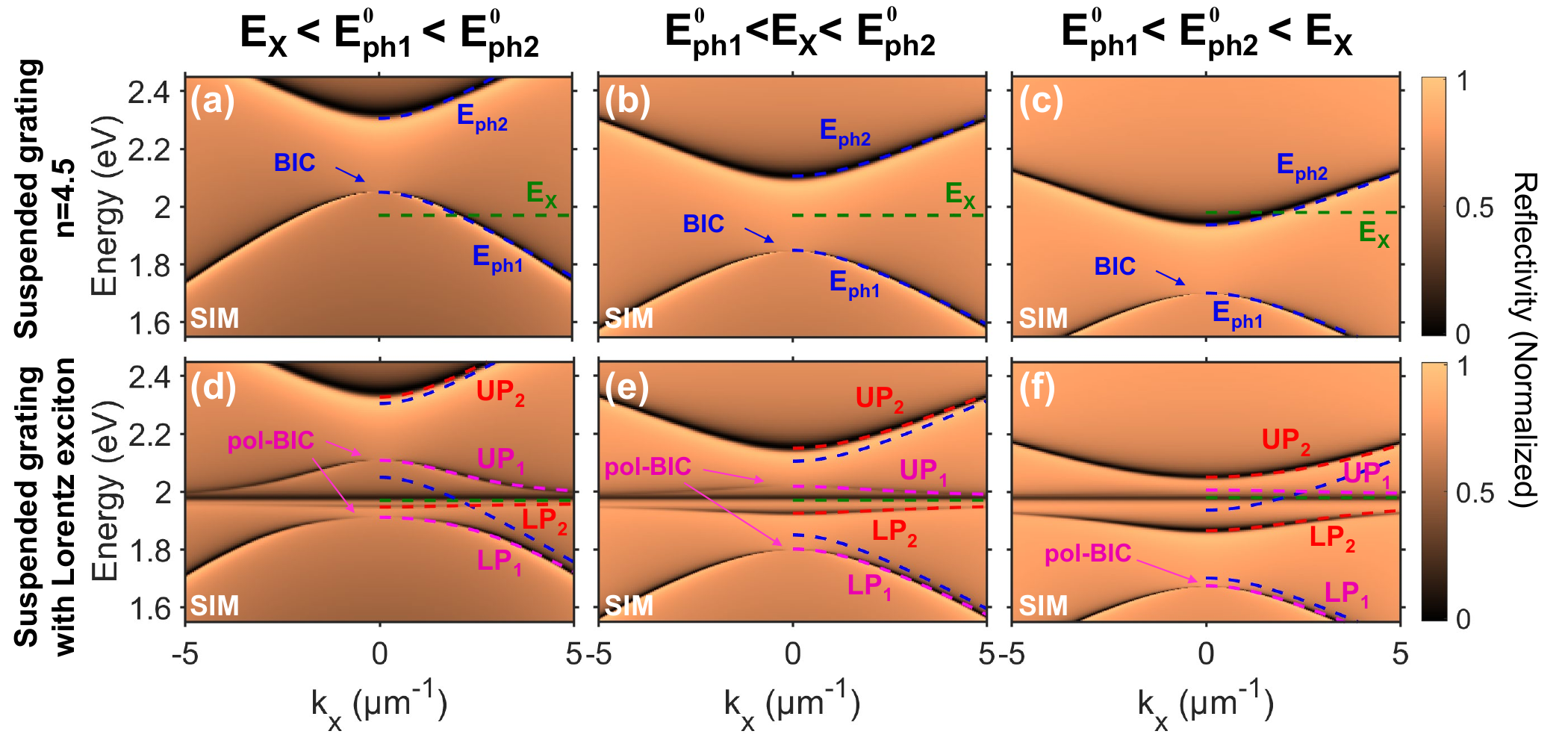}
\caption{\textbf{Numerical simulations of a double polaritonic system in the suspended gratings with different detunings.} (a-c), RCWA reflectivity simulations of the suspended grating with constant refractive index (n=4.5, k=0) shown in figure \ref{figure1}(a) for grating thickness of (a) 15nm, (b) 20nm, and (c) 25nm. The increase of the thickness redshifts the grating photonic modes $E_{ph1}$ and $E_{ph2}$ such that the excitonic energy $E_X$ lies below ($E_X<E_{ph1}^0<E_{ph2}^0$, (a)), in between ($E_{ph1}^0<E_X<E_{ph2}^0$, (b)), or above ($E_{ph1}^0<E_{ph2}^0<E_X$, (c)) the two photonic modes. BICs can be observed on the lower energetic photonic modes $E_{ph1}$. (d-f), RCWA reflectivity simulations of the suspended grating with an exciton-like refractive index shown in figure \ref{figure1}(b), with respectively the same grating thicknesses as in (a-c), and hence with same detunings between the grating photonic modes ($E_{ph1}$, $E_{ph2}$) and the exciton $E_X$. The strong coupling between the exciton and the grating photonic modes with different detunings lead to different double polaritonic systems described by the model in equation \ref{eq2}. The parameters used to fit the photonic and polaritonic modes are given in figure S3 of the supplementary. The first polaritonic set ($LP_{1}$, $UP_{1}$) are shown as pink dashed lines, and the second ($LP_{2}$, $UP_{2}$) as red dashed lines. The BICs in (a-c) also strongly couple with the exciton giving rise to polariton-BICs (pol-BICs) indicated in pink. Some excitons ($E_X$ green dashed lines) remain uncoupled to the photonic modes ($E_{ph1}$,$E_{ph2}$) and can be seen as horizontal lines in (d-f).}
 \label{figure2}
\end{figure*}

To understand the double exciton-polariton system, let us first consider the uncoupled photonic modes in a suspended grating (see figure \ref{figure1}(a)) of refractive index of 4.5 (close to the average refractive index of WS$_2$ at around 2 eV), and with a zero extinction coefficient. The grating consists of a single slab periodically etched along the x direction in wires extending in the y direction, and is characterized by its thickness, $t$, period $a$, and filling factor $FF$. The two uncoupled photonic modes, $E_{ph1}$ and $E_{ph2}$, of this structure result from the band folding and diffractive coupling of TE modes propagating along the x direction within the grating slab, and are described by (more details in section 1 of the supplementary):

\begin{equation}
	\label{eq1}
	E_{ph1,ph2}(k_x)=E_0\pm\sqrt{U^2+(v_g^{ph1,ph2}k_x)^2}
\end{equation} 

 where $U$ is the coupling strength of the diffractive coupling, $E_0$ the mid-gap energy, and $v_g^{ph1,ph2}$ the group velocity of each of the modes. As shown in figure \ref{figure1}(a), these two parabolic-like dispersions separated by a gap of $2U$ can be observed in RCWA (Rigorous Coupled Wave Analysis) reflectivity simulations (more details in methods) of a suspended grating ($a$=360nm, $FF$=0.8 $t$= 20nm, $n$=4.5), where $U$=255 meV, $E_0$=1.98 eV, $v_g^{ph1}$=72 meV.$\mu$m, $v_g^{ph2}$=62 meV.$\mu$m. Note that the low-energy photonic mode, $E_{ph1}$, vanishes when approaching zero momenta $k_x$. This dark state occurs due to destructive interferences induced by the in-plane symmetry of the photonic crystal and corresponds to a Bound State in the Continuum (BIC) \cite{VonNeumann1929,Friedrich1985}. 

Let us now consider a similar suspended grating with an excitonic material of refractive index $n_X$ modeled with a Lorentz oscillator (more details in methods). Figure \ref{figure1}(b) shows the sketch, refractive index, and RCWA reflectivity simulation of this suspended grating. In this structure, the two photonic modes strongly couple to the excitons leading to two sets of polaritonic dispersions ($LP_{1}$, $UP_{1}$) in pink, and ($LP_{2}$, $UP_{2}$) in red, with $LP$ for lower polariton, and $UP$ for upper polariton. The excitons interact with each grating mode individually because their mode volumes are spatially localized at different positions within the grating. This allows us to model this situation as the excitons coupling to each mode independently with the following four-level Hamiltonian from \cite{Sigurosson2024}: 

\vspace*{-5pt}

\begin{equation}
	\label{eq2}
	H_{polaritons}=
	\left(\begin{matrix} 
	E_{ph1}(k_x)  &   V      &    0    &     0     \\
            V  &    E_X     &    0    &     0     \\
              0  &    0       & E_{ph2}(k_x) &    V    \\
              0  &    0       &   V     &  E_X\\
 \end{matrix} \right)
\end{equation} \\

\begin{figure}[tp] 
   \includegraphics[width=\columnwidth]{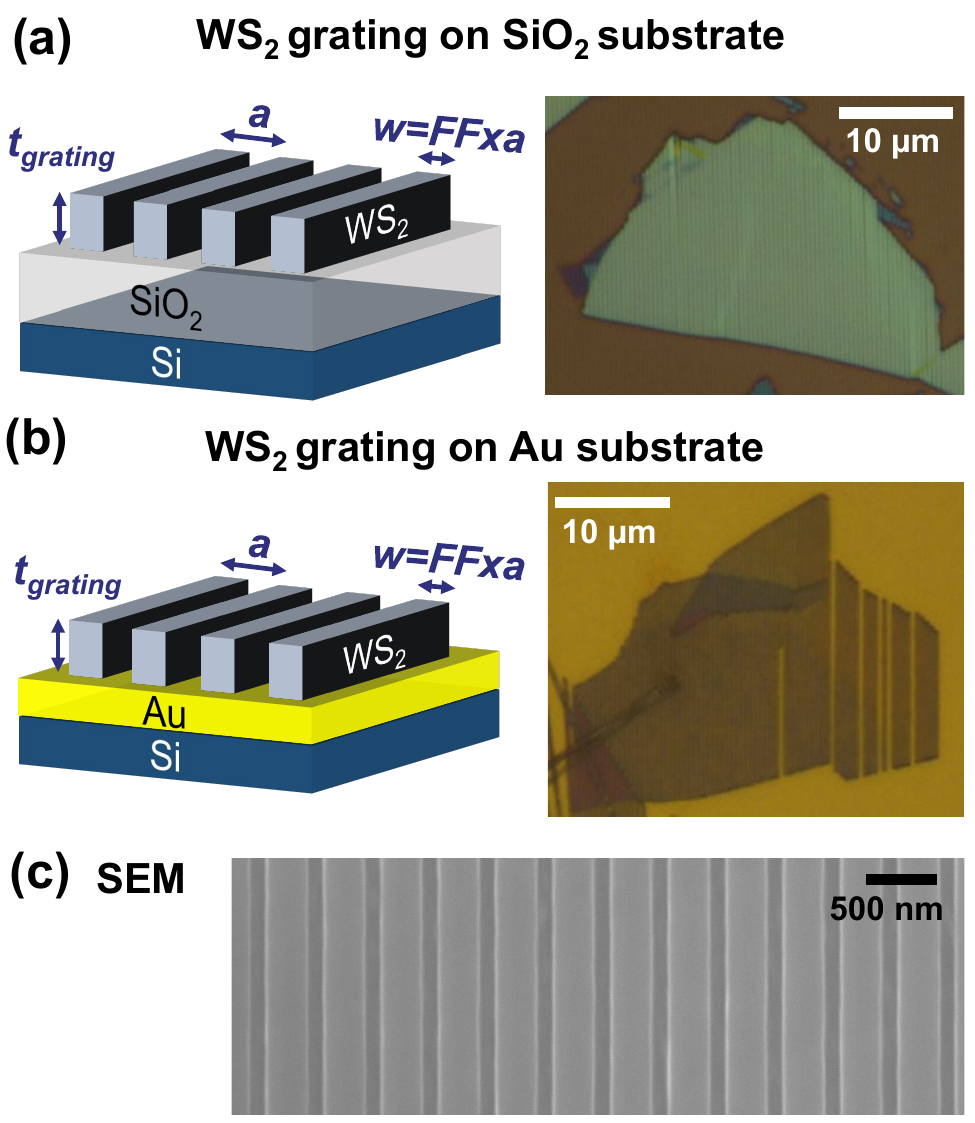}
  \caption{\textbf{Experimental WS$_2$-based 1D gratings.} (a-b) Sketch of the WS$_2$ gratings respectively on SiO$_2$/Si (a) and Au/Si (b) substrates along with a microscope images of two of the gratings. The period $a$ of these gratings are 350-400nm with filling factor $FF$ around 0.7 for the gratings on SiO$_2$/Si substrates and 0.55 on Au/Si substrates. (c) Scanning Electron Microscope (SEM) of one of the gratings on SiO$_2$/Si substrate. The Optical microscope images, AFM images and measured geometrical parameters of all the structures are shown in figure S5, S6 and S9 of the supplementary.}
 \label{figure3}
\end{figure}

\vspace*{-5pt}

where $E_{ph1}(k_x)$ and $E_{ph2}(k_x)$ are the two dispersions of the uncoupled photonic modes, $E_X$ the exciton energy, and $V$ the coupling strength between the exciton and the photonic modes (here $E_X$=1.98 eV, and $V$=90 meV). The exciton states which do not overlap with the photonic modes near-field distributions are not taken into account in this model as they do not contribute to the strong coupling but can be observed in the reflectivity simulations as a flat dispersion (in green). 

From figure \ref{figure1}(b), one can observe that the exciton $E_X$ is simultaneously strongly coupled with the bright state $E_{ph2}$ and the BIC seen in the photonic mode $E_{ph1}$, giving rise to two polariton-BICs (pol-BICs) \cite{Dang2022} observable on the polaritonic dispersions $LP_1$, and $UP_1$. 

To further study the double-polaritonic system, RCWA simulations were performed on the suspended gratings for different detunings, $\delta=E_0-E_X$, between the photonic modes and the exciton energy by changing the grating thickness. Figure \ref{figure2} presents the RCWA simulations of the suspended grating without an exciton ((a-c), $n$=4.5) and the excitonic suspended grating ((d-f), $n=n_X$) for grating thicknesses of 15, 20 and 25nm ($a$=360nm, $FF$=0.8) such that the excitonic energy is either below (a,d), in between (b,e), or above (c,f) the two uncoupled photonic modes. The parameters used to fit the photonic and polaritonic modes are given in figure S3 of the supplementary.

\begin{figure*}[ht!]
\centering
  \includegraphics[width=\linewidth]{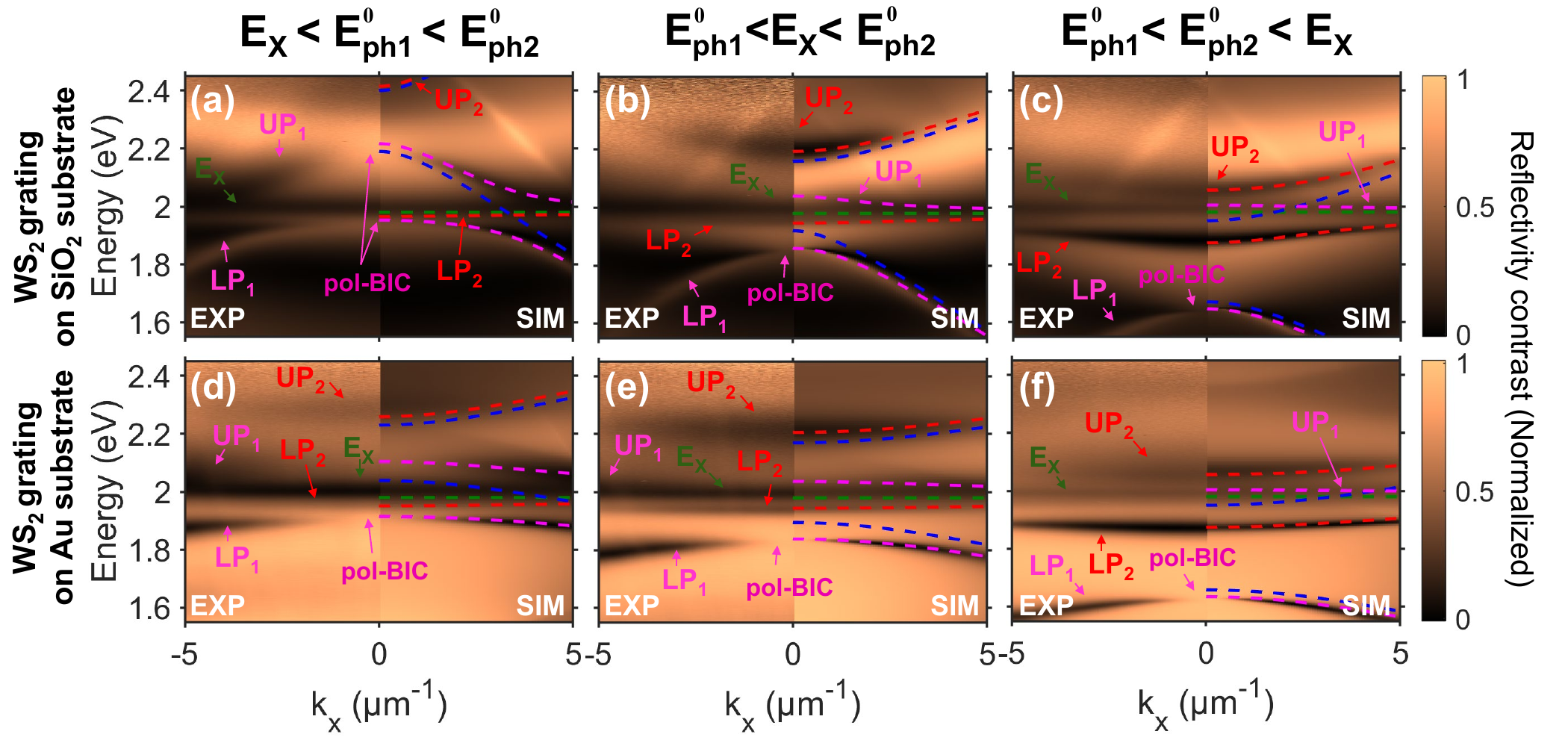}
  \caption{\textbf{Double polaritonic system in the WS$_2$ gratings on SiO$_2$ and Au substrate with different detunings.} Angle-resolved reflectivity contrast measurements (left panels) and RCWA (Rigorous Coupled Wave Analysis) numerical simulations (right panels) of the gratings on SiO$_2$/Si (a-c) and Au/Si (d-f) subtrates in the case when $E_X<E^0_{ph1}<E^0_{ph2}$ (a,d), $E^0_{ph1}<E_X<E^0_{ph2}$ (b,e), and $E^0_{ph1}<E^0_{ph2}<E_X$ (c,f). The experimental polaritonic and excitonic dispersions are fitted with the model in equation \ref{eq2} and the resulting dispersions, including the uncoupled photonic dispersions, are plotted on the right pannels for each structures. The photonic and polaritonic modes fitting parameters are given in figure S10 of the supplementary.  The photonic mode dispersions $E_{ph1}$ and $E_{ph2}$ are plotted as blue dashed line, the WS$_2$ exciton energy $E_X$ as a green dash line, the polaritonic dispersions $LP_{1(2)}$ and $UP_{1(2)}$ as pink(red) dashed lines, and the polariton-BICs (pol-BICs) are indicated in pink.} 
 \label{figure4}
\end{figure*}

When the exciton is below the two photonic modes in the exciton-less gratings ($E_X<E^0_{ph1}<E^0_{ph2}$ with $E^0_{ph1,ph2}=E_0\pm U$) in figure \ref{figure2}(a,d), the crossing between the exciton energy $E_X$ and the photonic mode $E_{ph1}$ leads to a well defined set of polaritons ($LP_1$,$UP_1$) with an anticrossing behaviour and a Rabi splitting, $\hbar\Omega=2V$, of 180 meV. The interaction between the exciton energy $E_X$ and the uncoupled photonic mode $E_{ph2}$ leads to a set of polaritons ($LP_2$,$UP_2$) much closer respectively to the uncoupled exciton energy and the uncoupled photonic mode $E_{ph2}$. Nevertheless, the coupling strength $V$ of 90 meV is still strong enough for one to discern the polariton dispersions. Moreover, the BIC mode on $E_{ph1}$ for the exciton-less suspended grating in figure \ref{figure2}(a) strongly couples with the exciton giving rise to two pol-BICs in the excitonic suspended grating in figure \ref{figure2}(d).

The case of the suspended gratings with an exciton above the two photonic modes ($E^0_{ph1}<E^0_{ph2}<E_X$) shown in figures \ref{figure2}(c,f), is similar to the previous case. The exciton state here is in resonance with the photonic mode $E_{ph2}$ and is above the photonic mode $E_{ph1}$. This leads to a well defined polariton set ($LP_2$,$UP_2$) with an anticrossing and a Rabi splitting, $\hbar\Omega=2V$, of 180 meV. However, in this case the upper polariton $UP_1$ is very close to the flat dispersion (in green) of the uncoupled excitons while having low contrast, which makes it difficult to be discerned.

Finally, when the exciton lies in the gap between the two photonic modes ($E^0_{ph1}<E_X<E^0_{ph2}$) in figures \ref{figure2}(b,e), the exciton line does not cross any of the two photonic modes. However, thanks to the high coupling strength $V$ of 90 meV, the two polariton sets ($LP_1$,$UP_1$) and ($LP_2$,$UP_2$) are well defined and far enough away from the uncoupled exciton flat dispersion to be easily resolved. Moreover, the BIC of the exciton-less suspended grating in \ref{figure2}(b,e) strongly couple with the exciton giving rise to two pol-BICs in the excitonic suspended grating figure \ref{figure2}(e).

\section{Experimental results}

The double-polaritonic system as a function of the detuning was then studied experimentally in six 1D WS$_2$-based gratings on SiO$_2$/Si and Au/Si substrates (see figure \ref{figure3}). The fabrication is detailed in the method section and the microscope and AFM images of the six structures are shown in figure S5 and S6 of the supplementary. Figure \ref{figure4} shows the angle-resolved reflectance contrast measurement (left panels) of the gratings using a spatial-filtering Fourier set-up (more details on methods), and their matching RCWA simulations (right panels) using the WS$_2$ refractive index from \cite{Zotev2023}. The period $a$ of these gratings were 400nm (350nm in figure \ref{figure4}(a)) and filling factor $FF$ around 0.7 for the grating on SiO$_2$/Si  substrates in figures \ref{figure4}(a-c) and 0.55 on Au/Si substrates in figures \ref{figure4}(d-f). The thicknesses of the grating are such that the WS$_2$ exciton energy either lies below (a, d), in between (b, d), or above (c, f) the two uncoupled photonic modes. The measured and simulated geometrical parameters of the gratings are given in figure S9, while the photonic and polaritonic modes fitting parameters are given in figure S10 of the supplementary. 

Just as for the numerical suspended grating, when the exciton lies below the two photonic modes ($E_X<E^0_{ph1}<E^0_{ph2}$) in figure \ref{figure4}(a,d), the polariton set ($LP_1$, $UP_1$) presents a well defined anticrossing with Rabi-splittings, $\hbar\Omega=2V$, of 160 (a) and 180 (d) meV. Moreover, one can simultaneously observe two pol-BICs on the lower and upper polaritons $LP_1$, and $UP_1$ in the case of the grating on the SiO$_2$ substrate in figure \ref{figure4}(a). When the exciton lies above the two photonic modes ($E^0_{ph1}<E^0_{ph2}<E_X$) in figures \ref{figure4}(c,f), it is the polariton set ($LP_2$, $UP_2$) which is well defined and Rabi splittings of 160 meV.

However, for both cases, due to the losses from the WS$_2$ absorption and the broader WS$_2$ exciton, the polariton modes are much broader. Consequently, the complementary polariton sets cannot be fully observed as one of the polariton dispersions ($LP_2$ in figures \ref{figure4}(a,d), $UP_1$ in figures \ref{figure4}(c,f)) are masked by the flat dispersion of the uncoupled excitons.

When the exciton lies within the gap between the two photonic modes ($E^0_{ph1}<E_X<E^0_{ph2}$) for the gratings on SiO$_2$ and Au subtrates in figures \ref{figure4}(b,e), one can easily observe three polaritonic dispersions: $LP_1$, $LP_2$, and $UP_2$ thanks to the large coupling strengths of 85 (b) and 90 (e) meV. However, unlike the simulated suspended grating in figure \ref{figure3}(e), the polariton dispersion $UP_1$ is difficult to be seen due to the WS$_2$ absorption and the broader uncoupled exciton flat dispersion. Indeed, the upper polariton dispersion is located in the tail of the uncoupled excitonic line.

\begin{figure*}[ht!]
\centering
  \includegraphics[width=\linewidth]{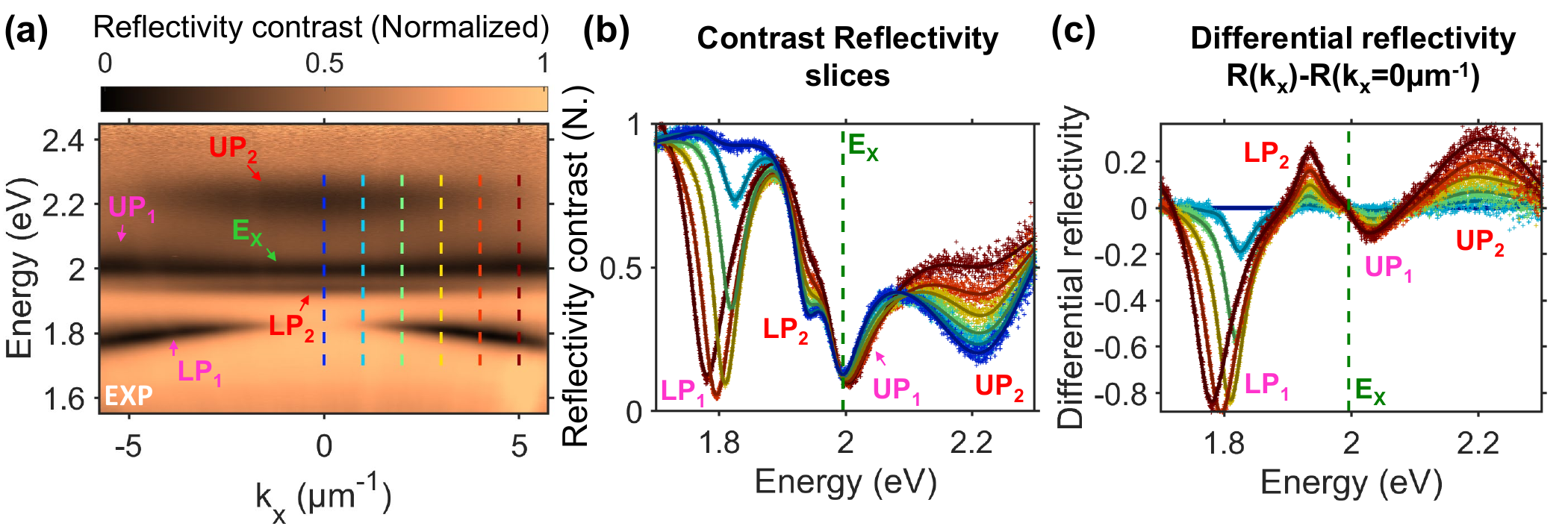}
  \caption{\textbf{Differential reflectivity slices to reveal the last polaritonic dispersion}. (a) Experimental reflectivity contrast of the WS$_2$ grating on Au substrate shown in figure \ref{figure4}(e) with vertical dashed lines corresponding to the slices further studied in (b,c). The different polaritonic dispersions $LP_1$, $LP_2$, and $UP_2$ are indicated with an arrow as well as the last polaritonic dispersion $UP_1$ located at the tail of the uncoupled exciton absorption. (b) Slices of the reflectivity map for wavevectors $k_x$ from 0 to 5 $\mu m^{-1}$ indicated by the dashed lines in (a). (c) Differential reflectivity of the slices compared to the slice at $k_x=0 \mu m^{-1}$ where the polaritonic dispersions $LP_1$, $LP_2$, and $UP_2$ are well defined. In these differential reflectivity slices, the uncoupled exciton contribution is removed due to the flatness of its dispersion, which reveals the previously unresolvable upper polariton $UP_1$.} 
 \label{figure5}
\end{figure*}

To reveal the unclear upper polariton $UP_1$ when the exciton is within the photonic gap ($E^0_{ph1}<E_X<E^0_{ph2}$), contrast reflectance spectra at different wavevector $k_x$ were further studied in the case of the gratings on Au substrate. The experimental contrast reflectance of figure \ref{figure4}(e) is reproduced in figure \ref{figure5}(a) with vertical dashed lines of different colours corresponding to the studied spectra. From these slices shown in figure \ref{figure5}(b), one can easily observe the reflectivity dips of the polariton dispersions $LP_1$, $LP_2$, and $UP_2$, as well as the uncoupled WS$_2$ exciton. However, one can observe that for large wavevectors $k_x$, the linewidth of the uncoupled exciton reflectivity dip broadens for energies higher than 2 eV, which corresponds to the hidden upper polariton $UP_1$ located in the tail of the uncoupled exciton. 

To confirm this hypothesis, we plot in figure \ref{figure5}(c) the difference between these spectra at different wavevectors and the spectrum at $k_x=0$ in order to remove the uncoupled exciton contribution. Indeed, the contribution of the flat uncoupled excitonic dispersion is constant over the change in wavevector $k_x$, and hence disappears in the differential spectra in figure \ref{figure5}(c). Similarly to the reflectivity spectra in figure \ref{figure5}(b), the polaritonic features $LP_1$, $LP_2$ and $UP_2$ are well defined in such differential reflectivity spectra in figure \ref{figure5}(c). More importantly, as the contribution of the uncoupled exciton is removed, one can observe an additional differential reflectivity dip for each spectrum corresponding to the previously unresolvable upper polariton $UP_1$. Consequently, the two sets of polaritons ($LP_1$,$UP_1$) and ($LP_2$,$UP_2$) can be experimentally observed simultaneously for the gratings in the case when the exciton lies within the grating photonic gap. 

The precense of the upper polariton $UP_1$ is further numerically confirmed for the structures shown in figures \ref{figure4}(b,e) by using their matching RCWA simulation. Indeed, as shown in figure S11 of the supplementary, the previously unresolvable upper polariton $UP_1$ can be observed from the RCWA simulations of the structure shown in figures \ref{figure4}(b,e), when the contrast reflectance signal is obtained when considering an uppaterned flake as a reference, which removes the contribution of the uncoupled excitons. Then, the numerical example of the suspended gratings, alongside the experimental results and matching RCWA simulations confirm the double polaritonic system in the WS$_2$-based gratings.

Finally, we note that the Rabi splitting reported here at room temperature are much larger than the ones from photonic structure based on III-V semiconductors with Rabi splittings in the order of a few to ten meV at cryogenic temperatures \cite{Hu2022,Gagel2024,StJean2017}, or the SiN and Ta$_2$O$_5$ gratings coupled respectively with WSe$_2$ \cite{Zhang2018} and MoSe$_2$ \cite{Kravtsov2020} monolayers with Rabi splitting around 20 meV also at cryogenic temperature. The Rabi splittings are, however, in the same order of magnitude than the ones reported in perovskite-based photonic crystals ranging from 150 meV to 210 meV \cite{Wu2024,Dang2022,Kim2021} and larger than the previously demonstrated polaritons in quasi-bulk WS$_2$ gratings with Rabi splitting of 100 meV \cite{Cho2023}, 120 meV \cite{Zong2021}, and 170 meV (without the plasmonic mode) \cite{Zhang2020}. However, in these WS$_2$ gratings, the strong coupling was only demonstrated with only one of the grating modes and polariton-BICs were not observed.

\vspace*{-10pt}

\section{Conclusion}

In conclusion, we have presented a highly tunable and versatile platform for the observation of a double polaritonic system based on the quasi-bulk van der Waals excitonic material. From 1D gratings made of the quasi-bulk WS$_2$, a double polaritonic system was demonstrated with large Rabi splittings ranging from 160 to 180 meV for different detunings of the system, in which the WS$_2$ excitons are simultaneously strongly coupled with the two grating photonic modes including the Bound State in the Continuum (BIC) of the lower energetic mode giving rise to polariton-BICs (pol-BICs). 

By using quasi-bulk TMDs, and combining the various advantages of 2D materials and polaritons, we believe that bulk 2D materials can open the way to a new prospect of future polaritonic photonic devices. Indeed, bulk 2D materials, used directly as photonic bulding blocks, have great properties to offer the fields of nanophotonics such as their large variety of transmission windows with low losses, high refractive indices, excitonic properties, or their compatibility with standard lithography and etching techniques. Moreover, the TMDs offer a versatility in choice of substrate such as demonstrated in this article. 

On the other hand, exciton-polaritons have also been attracting interest for photonic device applications as they inherit properties from both their light and matter counterparts. By combining bulk 2D materials and exciton-polaritons, a new direction in polaritonics can be instigated. Finally, using the transferability of TMDs, the highly tunable subwavelength gratings presented in our work could also be used in conjunction with slabs and whole photonic integrated circuits made from van der Waals or many other materials such as III-V membranes. Placing these gratings on top of such photonic structures could thus open the way to 3D nanophotonic heterointegration and straightforward fabrication of sophisticated polaritonic devices. 

\section*{Methods}

\textbf{Fabrication.} The WS$_2$ grating structures are fabricated on 290nm SiO2/Si substrates or on gold substrate (3nm Ti + 100nm Au on Si wafer) fabricated in National Graphene Institute (Manchester). The substrates are first cleaned by acetone (10 min) and isopropanol (IPA) (10 min) in an ultrasonic water bath then blow-dryed with nitrogen. Then, the substrates are treated with an oxygen plasma treatment to remove the residues and contaminants. Simultaneously, WS$_2$ flakes are mechanically exfoliated, repeatedly cleaved with a Nitto BT-130E-SL tape (low tack low residue tape), on commercially available WS$_2$ crystals (HQ Graphene, synthetic). Finally, the WS$_2$ flakes are transferred right away onto the clean substrates \cite{Huang2020}. Using optical microscopy, regions of the sample with high densities of exfoliated flakes are selected to be later etched.

For electron beam lithography (EBL), a 400 nm thick positive electron-beam resist CSAR 62 (AR-P 6200.13) is spin coated onto the samples. Each spin-coating procedures involve the film deposition and annealing process for solvent evaporation (bake at 180 degree for 2min). Then, the gratings structures are patterned into the resist layer using an EBL machine (Raith VOYAGER). 

Before the etching process, the samples are subsequently immersed in xylene, and IPA to dissolve the exposed areas of the positive resist CSAR 62, and get rid of chemical residues prior to blow-drying. The remaining patterned positive resist acts as a mask for the following reactive ion etching (RIE) of the WS$_2$ flakes. The chamber is first cleaned by Ar+H$_2$ and O$_2$ plasma, then a mixture of physical and chemical etching is processed using a combination of CHF$_3$ and SF$_6$ plasma. The residual resist film is finally removed by immersing the sample in hot 1165 resist remover (90$^\circ$C, 30 min) and hot acetone (90$^\circ$C, 30 min), followed by a rinsing of the sample with IPA and several seconds of O$_2$ ashing to remove the harder resist residues caused by RIE.

\textbf{RCWA Simulation} The RCWA simulations were performed using the $S^4$ package provided by Victor Liu and Shanhui Fan \cite{Liu2012}. In the case of the suspended gratings in figures \ref{figure1} and \ref{figure2}, the dielectric function of the grating slabs were either a constant refractive index, $n$=4.5, or obtained by the following Lorentz oscillator:

\begin{equation}
	\label{eq3}
	\epsilon(E)=n^2+\frac{A_X}{E_X^2-E^2+i\gamma_XE}
\end{equation} 

with $n=4.5$, $E_X=1.98eV$, $\gamma_X=10 meV$, and $A_X=0.8eV^2$. In the case of the RCWA simulations matching the experimental results in figure \ref{figure4}, the WS$_2$ layer refractive index was obtained from \cite{Zotev2023}. 

\textbf{Optical measurements.} Angle-resolved reflectivity contrast measurements shown in figures \ref{figure4} and \ref{figure5} were carried out using a spatial-filtering Fourier spectroscopy set-up (see figure S4 in SI). The sample is illuminated over a large area with an almost collimated white light using a 0.7 NA objective (100X Mitutoyo Plan Apo NIR) and a 150 mm lens placed in front of the objective. The collected reflected light by the objective is separated from the input signal with a beam splitter. The image of the sample area is then projected by the objective and a 250 mm lens onto a double slit that selects the desired rectangular region of the sample. A 600 mm lens performs the Fourier transform of the signal which is then projected with a 200 mm and 150 mm lenses onto the slit of a spectrometer selecting the wavevectors along the vertical direction. The diffractive grating inside the spectrometer disperses the light horizontally and the ($k_x$,$\lambda$) reflectivity dispersion signal from the rectangular region of interest is projected onto a 400x100 pixels CCD camera. 

\textbf{Data processing.} The reflectivity dispersions were both measured on the gratings and the substrate using the same double slit filter. The considered experimental reflectivty contrast signals were $R_{c}=(R_{g}-R_{ref})/R_{ref}$, with $R_g$ and $R_{ref}$ respectively the grating and substrate reflectivity. For the RCWA simulations matching the experimental results in figure \ref{figure4}, the reflectivity dispersions were both simulated on the structures and substrates and the same reflectivity contrasts signal were considered. For the RCWA simulated suspended grating in figures \ref{figure1} and \ref{figure2}, only the gratings reflectivities were considered. In the case of figure S11 in the supplementary information, the reference chosen for the reflectivity contrasts was an unpatterned $WS_2$ flake in order to remove the uncoupled exciton absorption on the reflectivity dispersions.

\section*{Acknowledgements}

All authors acknowledge the EPSRC grant EP/V026496/1

\section*{Supplemental Documents}

The following files are available free of charge.

\begin{itemize}
  \item Supplementary figures S1-S11 with grating photonic mode model, sketch of the spatial-filtering Fourier spectroscopy setup, additional microscope image of the gratings, AFM scans of the gratings, horizontal and vertical slices of the AFM scans, geometrical parameters measure din AFM and used in the RCWA simulations, parameters for the photonic and polaritonic dispersions, and additional simulation of the grating using bare flakes as reference.
\end{itemize}

\bibliography{mybib}

\end{document}


\title{Supplementary information for: \\ Simultaneous observation of bright and dark polariton states in subwavelength gratings made from quasi-bulk WS$_2$}

\author{Paul Bouteyre}
\email{p.bouteyre@sheffield.ac.uk}
\affiliation{Department of Physics and Astronomy, University of Sheffield, Sheffield S3 7RH, U.K} 
\author{Xuerong Hu}
\affiliation{Department of Physics and Astronomy, University of Sheffield, Sheffield S3 7RH, U.K} 
\author{Sam A. Randerson}
\affiliation{Department of Physics and Astronomy, University of Sheffield, Sheffield S3 7RH, U.K} 
\author{Panaiot G. Zotev}
\affiliation{Department of Physics and Astronomy, University of Sheffield, Sheffield S3 7RH, U.K} 
\author{Yue Wang}
\affiliation{School of Physics, Engineering and Technology, University of York, York, YO10 5DD, UK} 
\author{Alexander I. Tartakovskii}
\email{a.tartakovskii@sheffield.ac.uk}
\affiliation{Department of Physics and Astronomy, University of Sheffield, Sheffield S3 7RH, U.K} 

\date{\today}

\pacs{}

\maketitle

\vspace*{-30pt}

\section{Grating modes in a 1D subwavelength grating}


\subsection{Guided mode resonances in a single slab}

\vspace*{-5pt}

\begin{figure}[!htb]
\centering
\includegraphics[width=\textwidth]{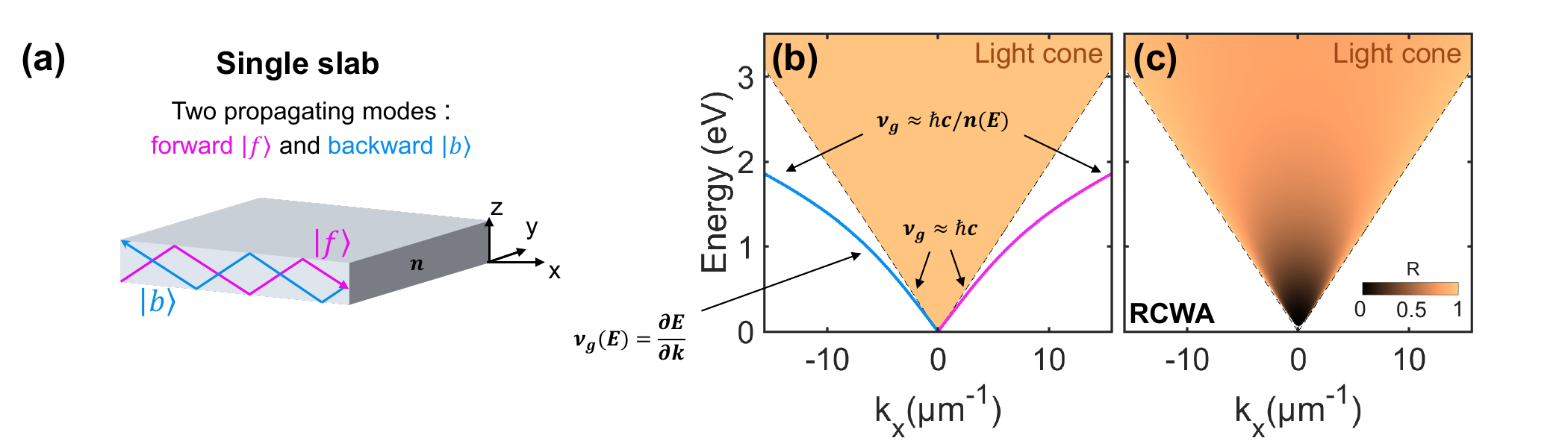}
    \caption{\textbf{Guided mode resonances in a single slab}. a) Sketch of a single slab with a backward and forward guided resonance. b) Dispersion of the backward and forward modes under the light cone. c) RCWA reflectivity simulation of a single slab of refractive index $n$=4.5 and thickness $t$=20nm. }
\label{figS1}
\end{figure}

In a suspended slab of refractive index $n(E)$ in air  (see figure \ref{figS1} (a)), forward and backward guided modes can propagate through total internal reflection within the slab. We consider here only their propagations along the x direction and that the slab is thin enough to support only one TE guided mode resonance (polarization along y) whose dispersion can be found in \cite{Huang2023}. As shown in figure \ref{figS1} (b), the guided modes dispersions of the backward and forward modes follow the light cone for low energies as the wavelengths are too long compared to the thickness of the slab. In this case, the group velocity of the mode, $\nu_g=\frac{\partial E}{\partial{k}} (eV.\mu m)$, is close to the one of the light line, i.e. $\nu_g\approx\hbar c$, with $\hbar$ the reduced Planck constant, and $c$ the light velocity. For higher energies, when the wavelength gets closer to the slab thickness, the group velocity approaches the group velocity of the light line in the slab, i.e. $\nu_g\approx\frac{\hbar c}{n(E)}$. More generally, the group velocity of the forward and backward modes is energy dependent, i.e. $v_g=v_g(E)$, both due to the dispersive nature of the guided resonance modes and the dispersive refractive index of the slab.

These guided mode dispersions shown in figure \ref{figS1} (b) are situated under the light cone, and therefore cannot be accessed from the far field perpendicular of the slab. For instance, these modes cannot be observed in the RCWA reflectivity simulation (see methods) of a single slab of refractive index $n$=4.5 and thickness $t$=20nm in figure \ref{figS1} (c).

\vspace*{-25pt}

\subsection{1D sub-wavelength grating photonic modes}

\vspace*{-5pt}

\begin{figure}[!htb]
\centering
\includegraphics[width=\textwidth]{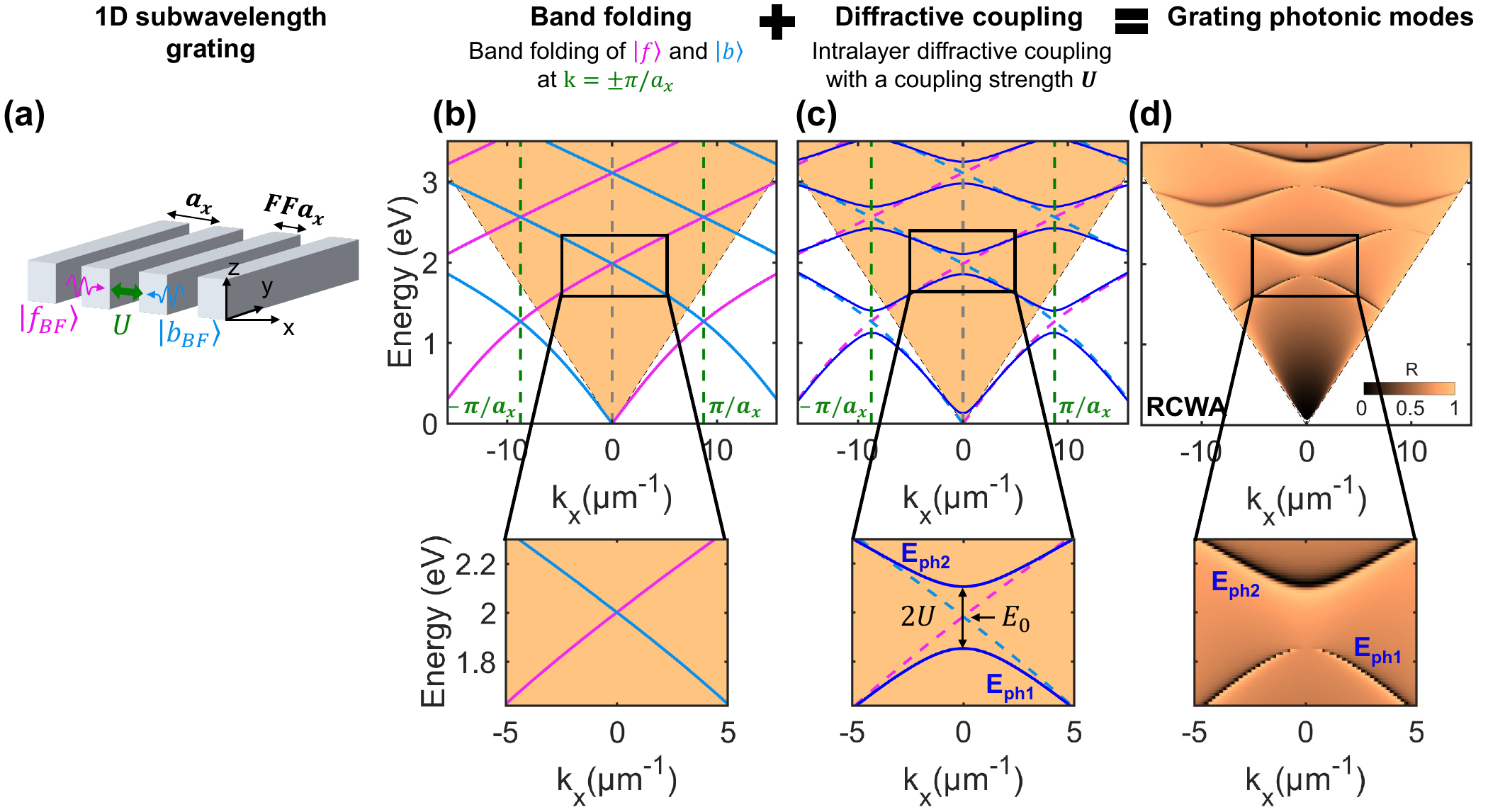}
    \caption{\textbf{1D sub-wavelength grating photonic mode}. (a) Sketch of the suspended 1D grating. (b) Band folding of the backward and forward modes at $k=\pm\pi/a$. (c) Diffractive coupling of the band folded modes. (d) RCWA reflectivity simulation of a 1D grating of refractive index $n$=4.5, thickness $t$=20nm, period $a$=360nm, and filling factor $FF$=0.8.}
\label{figS2}
\end{figure}

In the case of a sub-wavelength 1D grating (see figure \ref{figS2} (a)), the single slab detailed above is periodically etched along the x direction  with a period $a_x<\lambda$, which leads to a periodic square modulation of the slab permittivity $\epsilon(x)=n(x)^2$. From the model developed under perturbation theory in \cite{Sigurosson2024}, the grating photonic modes can be seen as the result of the diffractive coupling of the band-folded propagating modes over the Brillouin zone at $k_x=\pm\pi/a_x$ (see figures \ref{figS2} (b,c)). The considered propagating modes here are the ones that would propagate in a single slab of refractive index $n_{eff}$, effective refractive index of the periodically etched slab. Figure \ref{figS2} (d) shows the RCWA reflectivity simulation of a 1D grating of refractive index ($n$=4.5, $t$=20nm, $a$=360nm, $FF$=0.8), from which one can observe the grating modes above the light cone.

The photonic modes considered in the article are the first two modes at $\Gamma_0$ point (i.e. $k_xa_x/2\pi=0$) as shown in the square boxes in figures \ref{figS2} (c-d). Unlike the propagating modes in the single slab, these modes lie above the light cone and can be accessed from far field. These modes can be modeled with the following effective Hamiltonian \cite{Sigurosson2024}:

\begin{equation}
	\label{eqS1}
	H_{grating}=
	\left(\begin{matrix} 
	E_{0}+\nu_g(E)k_x  &   U   \\
            U          &   E_{0}-\nu_g(E)k_x        \\
 \end{matrix} \right)
\end{equation} \\

describing the diffractive coupling, with coupling strength $U$, of the first two band-folded backward and forward modes of dispersion $E_0\pm\nu_g(E)k_x$, with $E_0$ the energy of the modes crossing. In here, the losses and coupling through the radiative channels are ignored. After diagonalization of the Hamiltonian, one can retrieve the grating photonic modes:

\begin{equation}
	\label{eqS2}
	E_{ph1,ph2}=E_0\pm\sqrt{U^2+(\nu_g(E)k_x)^2},
\end{equation} \\

which depends on the dispersive velocity group $\nu_g(E)$. For the sake of simplicity, we consider in this article a given velocity group for a given grating photonic mode $\nu_g^{ph1}$, and $\nu_g^{ph2}$ such that:

\begin{equation}
	\label{eqS3}
	E_{ph1,ph2}=E_0\pm\sqrt{U^2+(\nu_g^{ph1,ph2}k_x)^2}.
\end{equation} \\

\newpage
\section{Supplementary figures}

\begin{figure}[!htb]
\centering
\includegraphics[width=.95\textwidth]{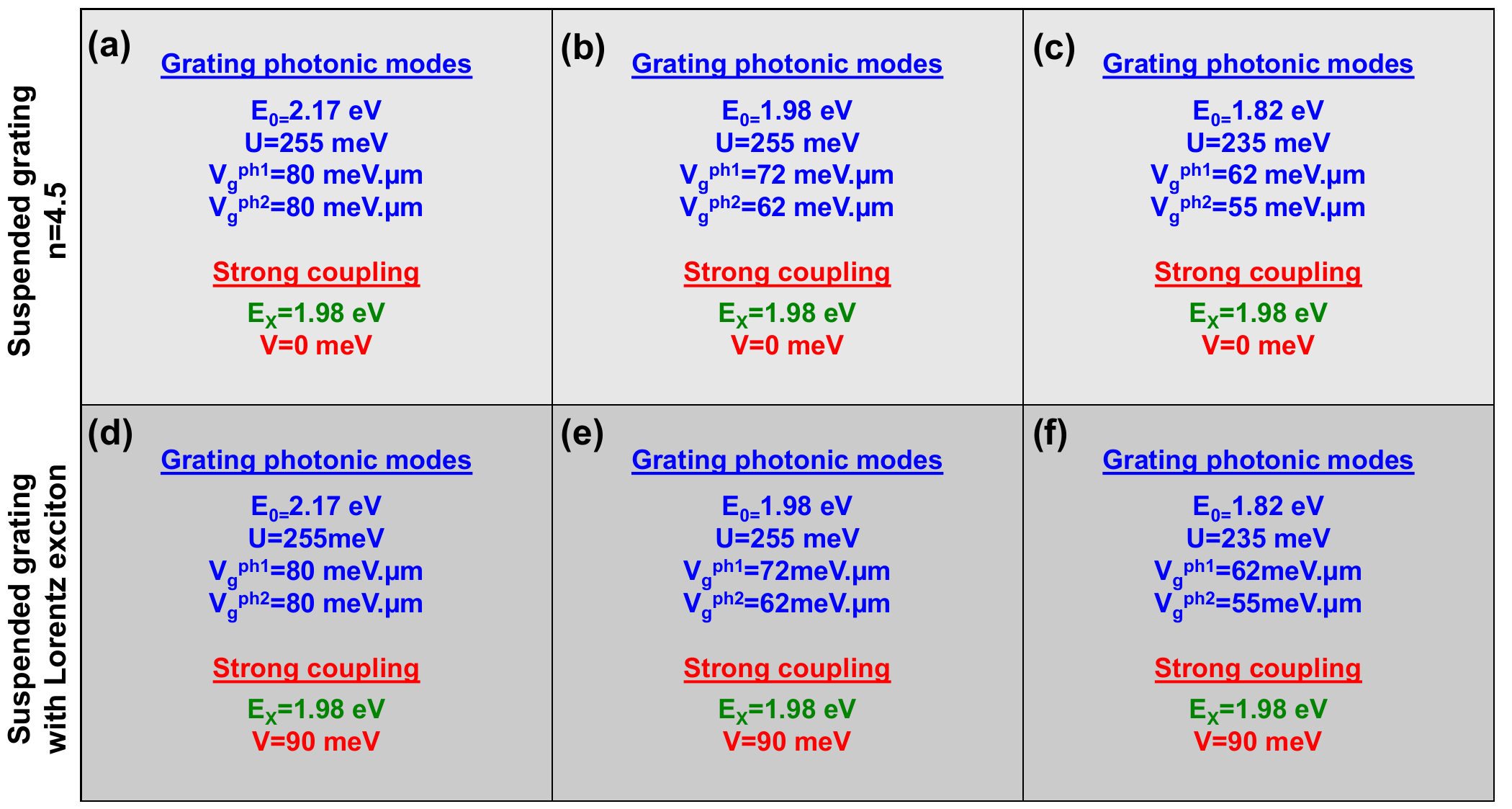}
    \caption{Fitting parameters used to fit the photonic and polaritonic modes of figure 2 of the article. (a-f) correspond respectively to the parameters used for the six suspended gratings presented in figure 2 (a-f) of the article ((a-c) for $n=4.5$, (d-f) for $n=n_X$, with $n_X$ a Lorentz oscillator (see methods) using $E_X=1.98eV$, $\gamma_X=10 meV$, and $A_X=0.8eV^2$).}
\label{figS3}
\end{figure}

\begin{figure}[!htb]
\centering
\includegraphics[width=.95\textwidth]{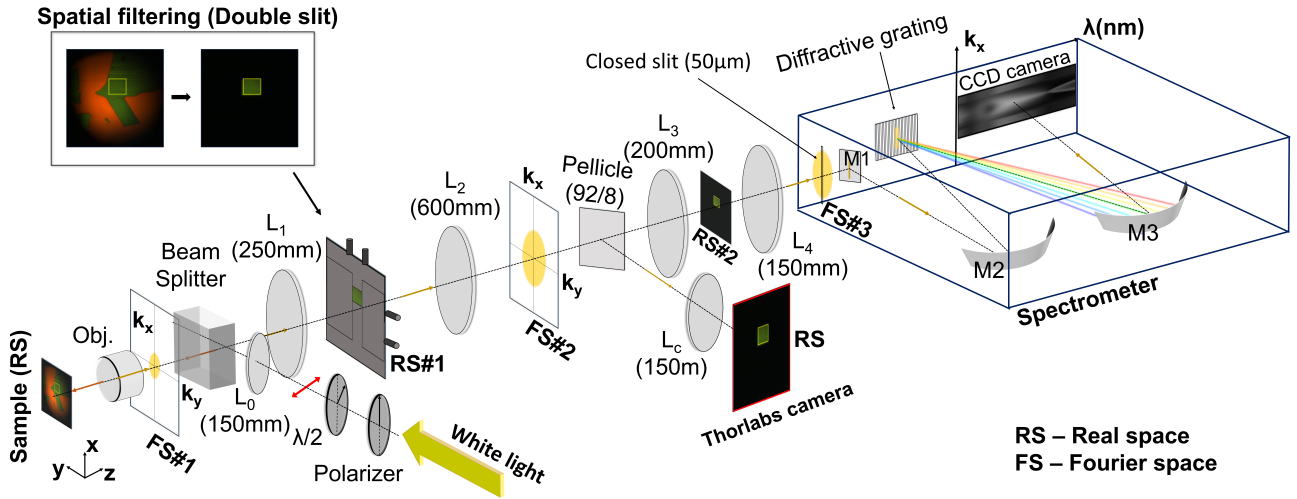}
    \caption{Angle-resolved reflectivity contrast measurements were carried out using a spatial-filtering Fourier set-up. The sample is illuminated over a large area with an almost collimated white light using a 0.7 NA objective (100X Mitutoyo Plan Apo NIR) and a 150 mm lens ($L_0$) before the objective. The collected reflected light by the objective is separated from the input signal with a beam splitter. The image of the sample is then projected by the objective and a 250 mm lens ($L_1$) onto a double slit which selects the desired rectangular region of the sample. A 600 mm lens ($L_2$) placed at focal length behind the spatially-filtered real space performs the Fourier transform of the signal, which is then projected with a set of two lenses (200 mm ($L_3$), 150 mm ($L_4$)) onto the slit of a spectrometer which selects the wavevectors along the vertical direction. The diffractive grating inside the spectrometer disperses the light horizontally and the signal is projected onto a 1340x400 pixels CCD camera resulting in a ($k_x$,$\lambda$) reflectivity dispersion signal from the rectangular region of interest.}
\label{figS4}
\end{figure}

\begin{figure}[!htb]
\centering
\includegraphics[width=.94\textwidth]{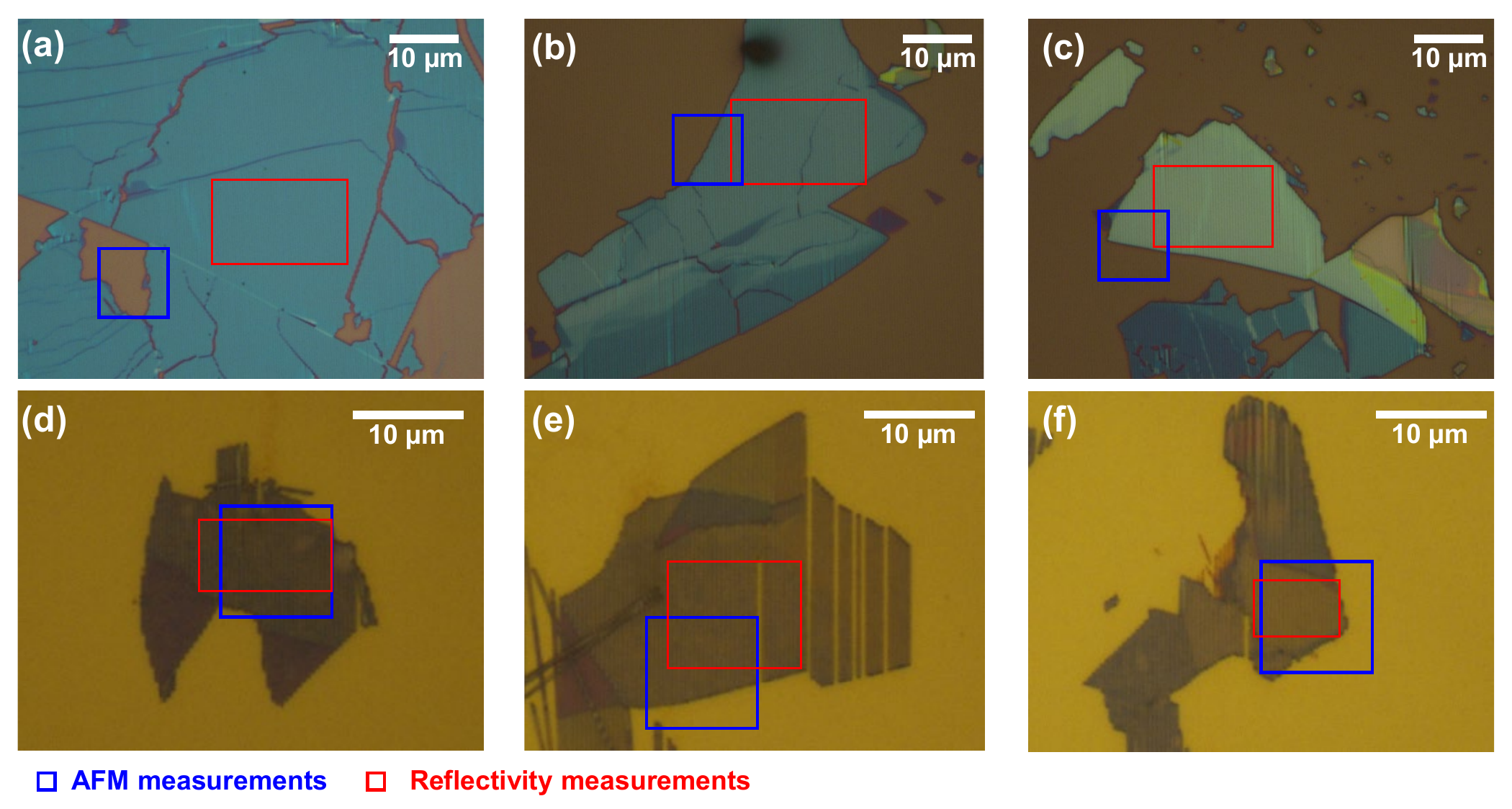}
    \caption{Optical microscope images of the fabricated WS$_2$ gratings. (a-f) correspond respectively to the six gratings presented in figure 4 (a-f) of the article ((a-c) for the WS$_2$ gratings on SiO$_2$/Si substrate, (d-f) on Au/Si substrate). The blue boxes correspond to the area of the AFM scan in figure \ref{figS6} and the red boxes to the areas of the reflectivity measurement using the set-up presented in fig \ref{figS4}.}
\label{figS5}
\end{figure}

\begin{figure}[!htb]
\centering
\includegraphics[width=.94\textwidth]{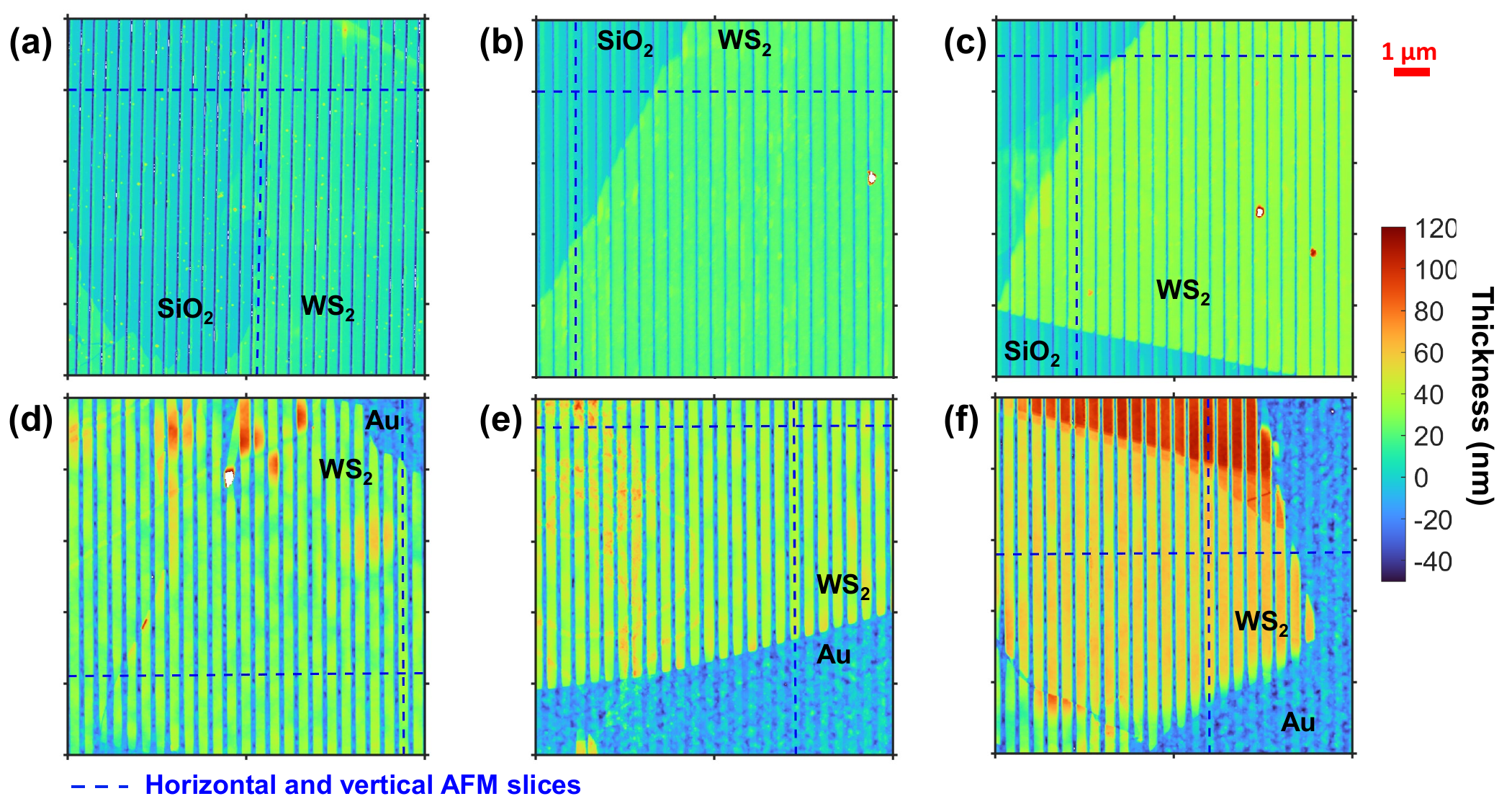}
    \caption{AFM images of the fabricated WS$_2$ gratings. (a-f) correspond respectively to the six gratings presented in figure 4 (a-f) of the article ((a-c) for the WS$_2$ gratings on SiO$_2$/Si substrate, (d-f) on Au/Si substrate). The areas of the AFM scans are indicated by the blue boxes in fig \ref{figS5}. The dashed vertical and horizontal blue lines correspond to the AFM slices presented in fig \ref{figS7} and \ref{figS8} for the determination of the geometrical of the fabricated gratings (period, thickness, and filling factors). The colormap was chosen to show the different thicknesses of the grating while showing the etching of the substrates.}
\label{figS6}
\end{figure}

\begin{figure}[!htb]
\centering
\includegraphics[width=.97\textwidth]{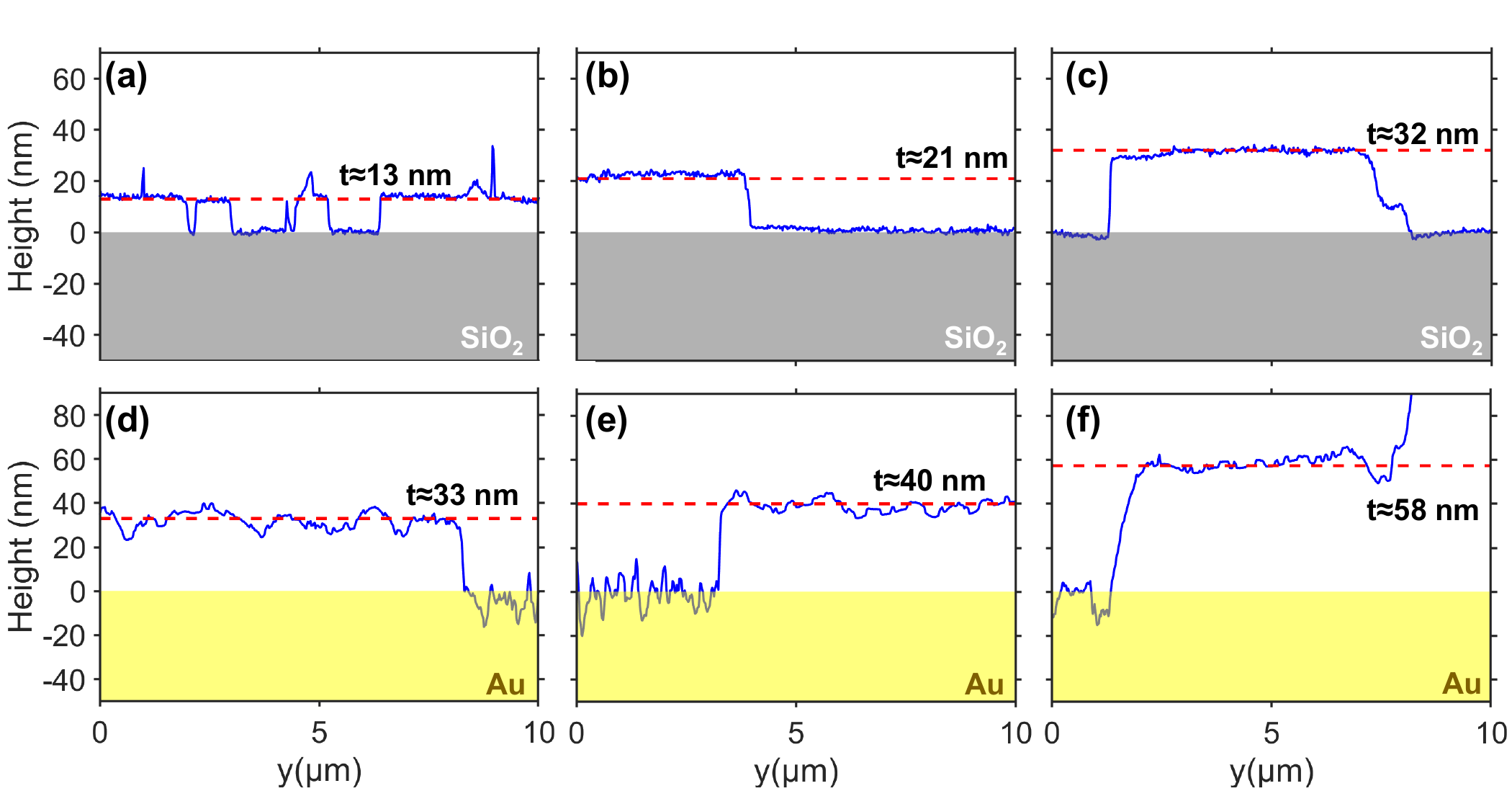}
    \caption{Vertical slices of the AFM images indicated by vertical dashed blue lines in fig \ref{figS6} to get the grating thicknesses. (a-f) correspond respectively to the six gratings presented in figure 4 (a-f) of the article ((a-c) for the WS$_2$ gratings on SiO$_2$/Si substrate, (d-f) on Au/Si substrate). The geometrical parameters of the fabricated grating are summarized in figure \ref{figS9} along with the RCWA parameters.}
\label{figS7}
\end{figure}

\begin{figure}[!htb]
\centering
\includegraphics[width=.97\textwidth]{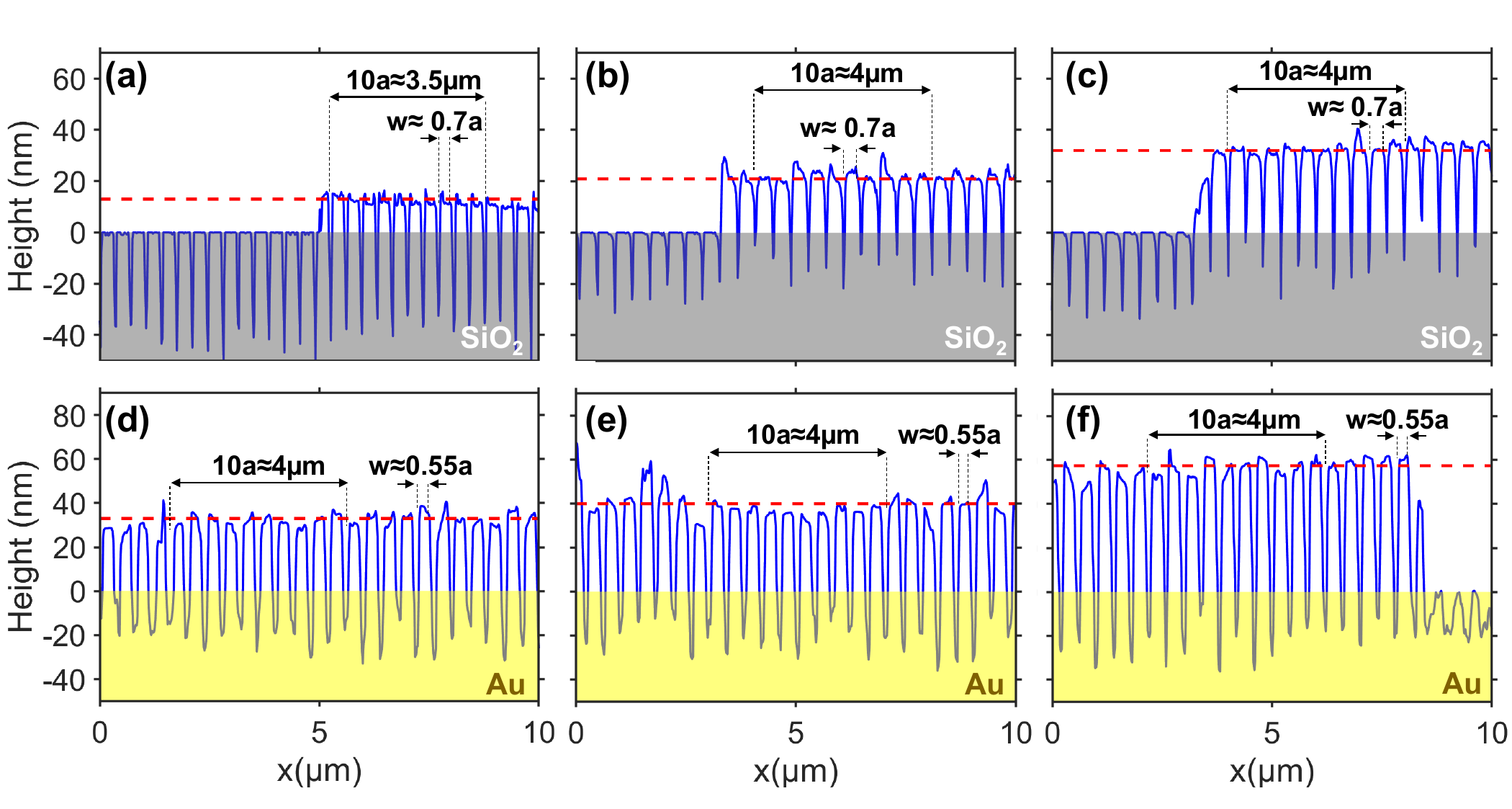}
    \caption{Horizontal slices of the AFM images indicated by horizontal dashed blue lines in fig \ref{figS6} to get the grating period, filling factor and substrate etch thicknesses. (a-f) correspond respectively to the six gratings presented in figure 4 (a-f) of the article ((a-c) for the WS$_2$ gratings on SiO$_2$/Si substrate, (d-f) on Au/Si substrate). The red dash lines correspond to the grating thicknesses obtain in the vertical slices in figure \ref{figS7}. The geometrical parameters of the fabricated grating are summarized in figure \ref{figS9} along with the RCWA parameters.}
\label{figS8}
\end{figure}

\begin{figure}[!htb]
\centering
\includegraphics[width=\textwidth]{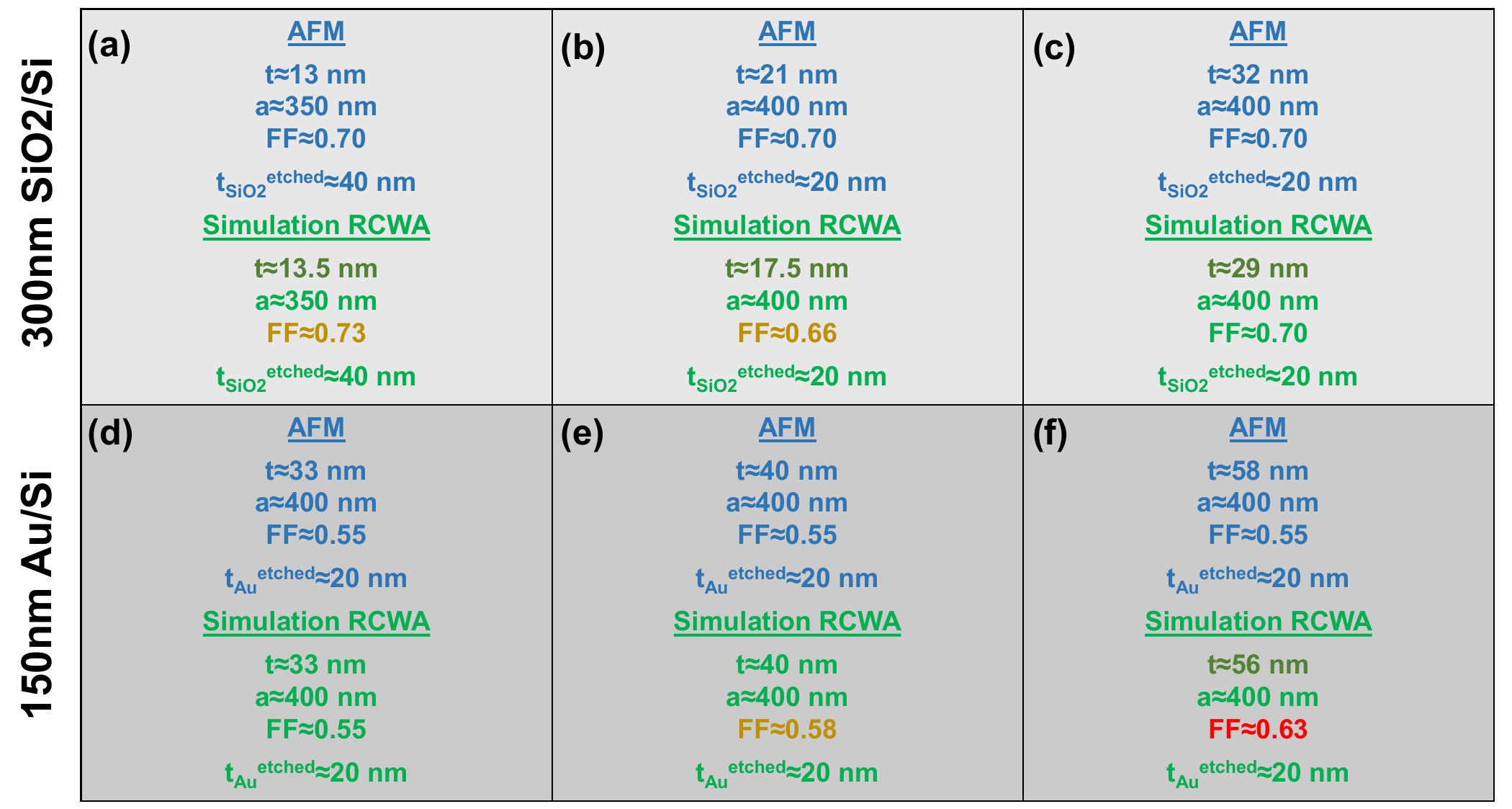}
    \caption{Summary of the measured geometrical parameters of the fabricated gratings from the AFM scans in figure \ref{figS6}, \ref{figS7}, and \ref{figS8} along with the RCWA parameters used to match the experimental reflectivity dispersion in figure4 of the article. (a-f) correspond respectively to the six gratings presented in figure 4 (a-f) of the article ((a-c) for the WS$_2$ gratings on SiO$_2$/Si substrate, (d-f) on Au/Si substrate). For the RCWA parameters, the green font indicates a really good match between the RCWA parameters and AFM measurements, the orange and red font indicate a moderate and large deviation between the two.}
\label{figS9}
\end{figure}

\begin{figure}[!htb]
\centering
\includegraphics[width=\textwidth]{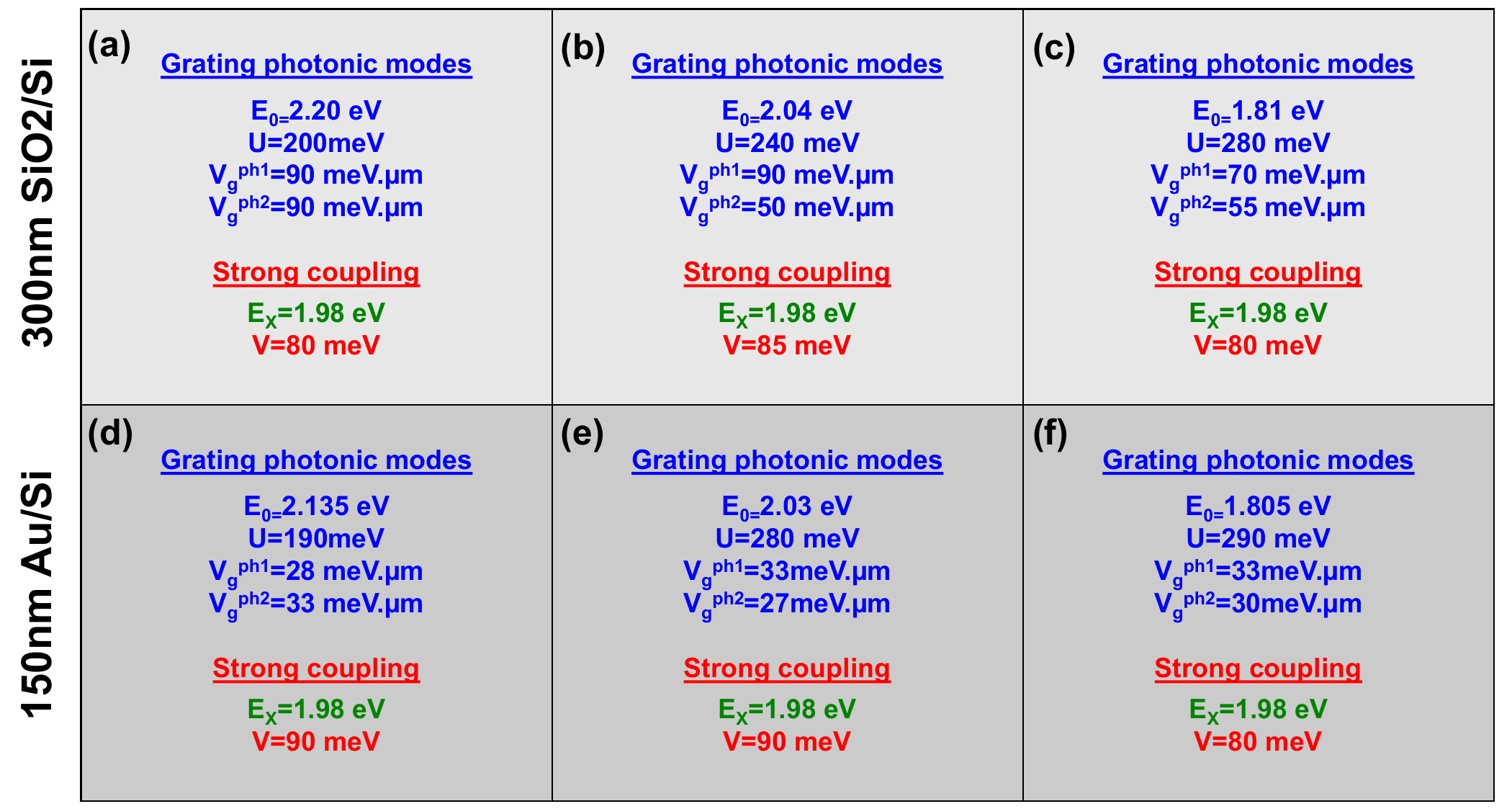}
    \caption{Fitting parameters used to fit the photonic and polaritonic modes of figure 4 of the article. (a-f) correspond respectively to the parameters used for the six suspended gratings presented in figure 4 (a-f) of the article ((a-c) for the WS$_2$ gratings on SiO$_2$/Si substrate, (d-f) on Au/Si substrate).}
\label{figS10}
\end{figure}

\begin{figure}[!htb]
\centering
\includegraphics[width=\textwidth]{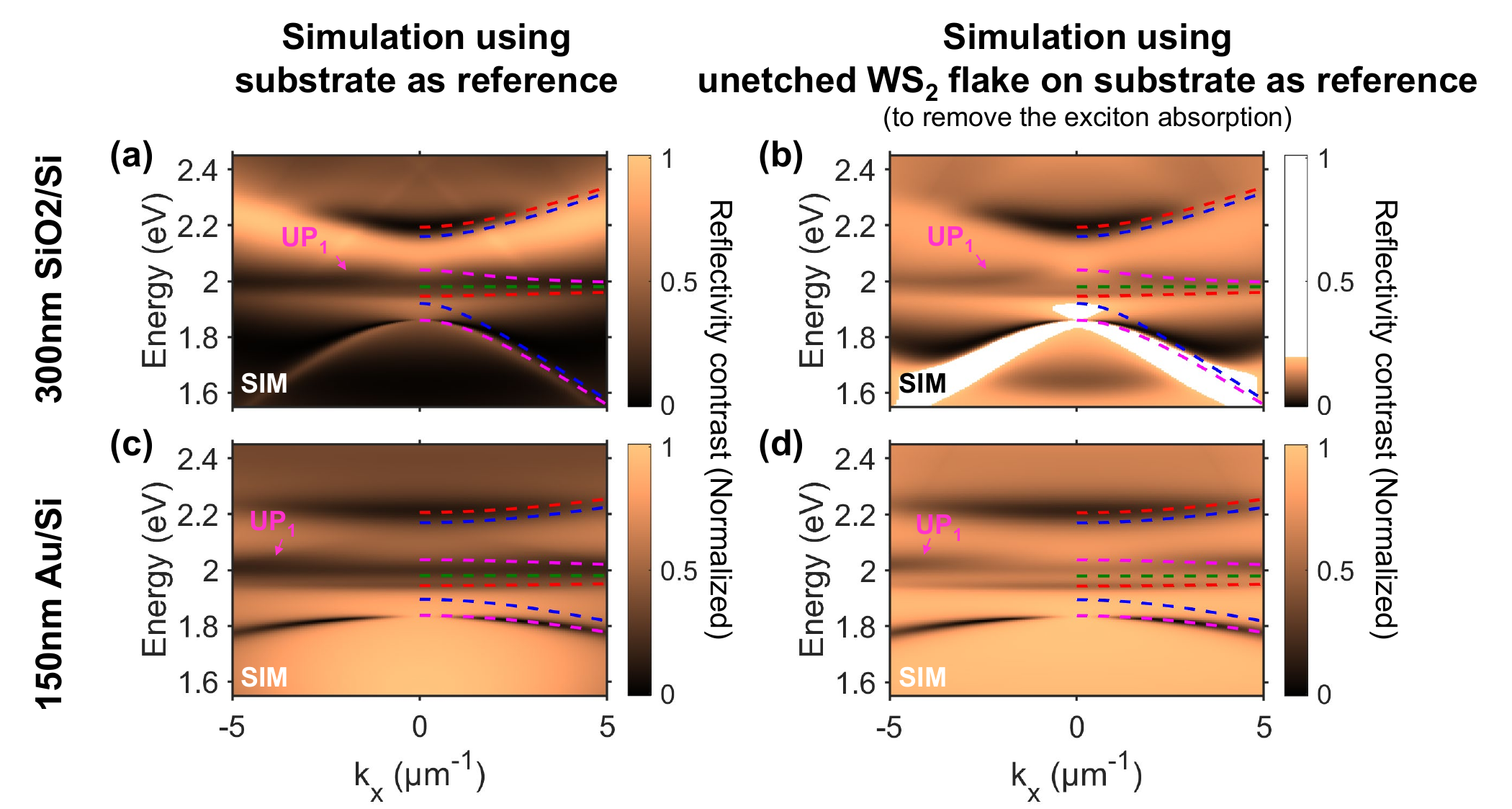}
    \caption{Angle-resolved reflectivity contrast simulations shown in figure 4 (b) and (e) of the article in the case when the exciton lies within the gap of the grating photonic modes (i.e. $E^0_{ph1}<E_X<E^0_{ph1}$). (a) and (c) are respectively the same RCWA reflectivity contrasts shown on the right panels of figure 4 (b) and (e). (b) and (d) are the same RCWA relfectivity contrast using an unpatterned flake on substrate as a reference (more details in the method section of the article). With the unpatterned flake as reference, the simulated reflectivity dispersions reveals the missing upper polariton dispersion $UP_1$ which could only barely be seen when the substrate was used as reference. }
\label{figS11}
\end{figure}

\FloatBarrier
\bibliography{mybib}